\documentclass[aps,twocolumn,nofootinbib,preprintnumbers,superscriptaddress]{revtex4}

\usepackage{amsmath,amssymb}
\usepackage{graphicx}
\usepackage{slashed} 

\begin{document}

\title
{
Fermionic probes of local quantum criticality in one dimension
}
\preprint{DCPT-11/27}

\author{Mukund Rangamani}
\author{Benjamin Withers}
\affiliation{Centre for Particle Theory \& Department of
Mathematical Sciences, Science Laboratories, South Road, Durham DH1 3LE, United Kingdom\\
\tt{mukund.rangamani, b.s.withers@durham.ac.uk}}

	\date{June 2011}

\begin{abstract}
We study the spectral functions of fermionic operators in $1+1$ dimensional  $SU(N)$ Super Yang-Mills theory with 16 supercharges at finite density using the holographically dual D1-brane geometry. This system exhibits quasi-particle peaks indicating the existence of Fermi points about which excitations have a finite relaxation time, in contrast with the Tomonaga-Luttinger model. The finite width may be attributed to the non-trivial interactions of the probe operators with the background density matrix, modeled holographically as a charged black hole. We show that the fermionic correlators can in fact be deduced from known results for fermion probes of  the charged AdS$_4$ black hole background, owing to some remarkable coincidences in supergravity truncations.
\end{abstract}


\maketitle

\section{Introduction}
Interacting Fermi systems in one spatial dimension (1d) provide the textbook example of departure from conventional Fermi liquid theory. For reasons associated with the restricted phase space in 1d, particle-hole pairs are indistinguishable from collective excitations. Whilst in higher dimensional fermionic systems have low energy fermionic quasiparticles, in 1d we have bosonic particles -- spin and charge density waves. 

The  prototype model for interacting 1d fermions is the Tomonaga-Luttinger (TL) model which involves linearising the single particle dispersion about the Fermi points. This has the advantage of being exactly soluble via bosonisation \cite{Haldane:1981fk} and is a conformal field theory. Retaining nonlinearity in the spectrum has recently shown to be possible and the spectral function for nonlinear Luttinger (NLL) liquids has recently been computed in perturbation theory \cite{Khodas:2007uq,Imambekov:2009fk}.

In this paper we are going to explore the dynamics of fermions in a non-conformal field theory, $SU(N)$ Super Yang-Mills with 16 supercharges in $1+1$ dimensions (SYM$_2$)  at finite density. We will extract the spectral function $A(\omega,k)$ for this system in the planar limit $N \to \infty$ by exploiting the fact that it admits a holographic dual \cite{Itzhaki:1998dd} for intermediate energy scales, $E$,  $\sqrt{\lambda}  \geq  E \geq \sqrt{\lambda}\, N^{-\frac{1}{3}}$, in terms of the near-horizon geometry of a stack of $N \gg 1$ parallel D1-branes in Type IIB supergravity. Here we introduced the 't Hooft coupling $\lambda \equiv g_{\text{YM}}^2 N$ whose engineering dimension is $2$. 

As SYM$_2$ is a non-conformal theory, one anticipates that the spectral function will deviate significantly from the TL form.\footnote{Luttinger liquids have been modelled holographically in terms of fermionic fields probing a BTZ black hole \cite{Balasubramanian:2010sc}.} In the latter one encounters power law singularities with exponents set by the interaction strength. Explicit breaking of scaling symmetry ought to soften this behaviour and indeed we find similarities with the NLL predictions. What is however a lot more curious is the connection we uncover between this D1 system and the analysis of non-Fermi liquids from holography undertaken in \cite{Liu:2009dm,Faulkner:2009wj,Faulkner:2011tm} (see also \cite{Lee:2008xf,Cubrovic:2009ye}) for 2d systems. These set-ups are now well understood in the semi-holographic framework \cite{Faulkner:2010tq}. In these setups deviation from Fermi liquid behaviour is attributed to  local quantum criticality arising from the near-horizon AdS$_2$ region encountered in extremal black hole geometries. 

While the example we focus on also has a local quantum critical region, we shall in fact find that the effective holographic master field equations that govern the spectral function are identical despite the seemingly different microscopic nature of the field theories. To be precise we find that the spectral functions for fermionic operators in SYM$_2$ are  identical to those obtained in the $2+1$ dimensional field theory of M2-branes. This is made possible by some remarkable coincidences in supergravity as we shall see in due course.

\section{The gravitational set-up}

SYM$_2$ for $E \in \sqrt{\lambda} \, (N^{-\frac{1}{3}}, 1)$ is described holographically in terms of the supergravity geometry obtained by taking the near-horizon limit of a stack of $N$ D1-branes. As we are interested in the dynamics of the theory in the grand canonical ensemble with non-zero temperature and charge density, we consider configurations where the D1-branes are non-extremal and spinning in the transverse ${\mathbb R}^8 \subset {\mathbb R}^{9,1}$. This turns on $R$-charges along the Cartan $U(1)^4 \subset SO(8)$ and for simplicity we focus on the equally-spinning case when these are all equal.
 
Whilst the $9+1$ dimensional ($10$D) Type IIB solution for this background configuration is easily written down \cite{Harmark:1999xt}, it is convenient to work with a KK-reduced solution in 3d which is obtained by reducing Type IIB equations of motion on a ${\bf S}^7$ with internal angular momentum \cite{David:2010qc}. This Lagrangian comprises of a dilaton $\phi$,  a gauge potential, $A$ and the metric, with the effective action 
\begin{equation}\label{s3}
S_3=  \frac{1}{16\pi\,G_3}  \int d^3x\, \sqrt{-g} \left(R - 2 (\partial \phi)^2 - e^{-2\phi} F^2 + 6 \frac{e^{2\phi}}{L^2}\right)
\end{equation}
where $F=dA$. The spinning D1-brane is described as a charged, dilatonic black hole solution of \eqref{s3}. Employing a prescient radial parametrisation given by the dilaton, $r \equiv e^{-\phi}$, it is given by
\begin{eqnarray}\label{ds3}
ds_3^2 &=& r^4\left(-f(r) dt^2 + dx^2\right) + \frac{L^2}{f(r)} dr^2\\
A &=& Q \, \left(1- \frac{1}{r}\right) dt
\label{a3}
\end{eqnarray}
where
\begin{equation}\label{f3}
f(r) = 1 + \frac{Q^2}{r^4}- \frac{1+Q^2}{r^3}.
\end{equation}
The horizon is located at $r=1$ and the black hole and dual field theory temperature is given by $T=\frac{1}{4\pi L} (3-Q^2)$. Alert readers will recognise the close similarity between this solution and the four-dimensional Reissner-Nordstrom AdS solution, which will have an important role to play in the following.

\subsection{Fermionic probes}

The field theory SYM$_2$ has fermionic operators $\mathcal{O}_\Psi$ residing in super-multiplets whose lowest component comprises of scalar chiral primaries, ${\rm Tr}\left(X^{(i_1} \cdots X^{i_n)}\right)$, transforming symmetric traceless representations of the $SO(8)$ R-symmetry group. We recall that the Lagrangian for SYM$_2$ can be obtained by dimensionally reducing the 10d ${\cal N} =1$ SYM Lagrangian on a ${\bf T}^8$ and hence apart from the gluons it contains adjoint scalars $X^i$ which transform as vectors under $SO(8)$. 

The dimensions of the operators $\mathcal{O}_\Psi$ can be computed by examining the two-point function in the vacuum. In the energy regime where the dual supergravity description is valid these take on integral values $\Delta_\Psi = n+3$ with $n \ge 0$. Equivalently, this spectrum can be described in terms of fermionic supergravity fields propagating in effective 3d D1-brane geometry by studying linearised KK fluctuations about the vacuum D1-brane solution. A comprehensive analysis of this reduction was carried out in \cite{Morales:2002ys} and it was found that the masses are related to their conformal dimension via
\begin{equation}
\Delta_\Psi = d_\text{eff} + |m L| \ , \qquad d_\text{eff} = 3 . 
\label{}
\end{equation}	

To compute the spectral function of $\mathcal{O}_\Psi$ using supergravity it suffices to consider a quadratic fermion effective action. For a fermionic mode $\Psi$ with mass $m$,
we use
\begin{equation}
S_\Psi = \int d^3x \sqrt{-g}\, i \, \left(\bar{\Psi} \slashed D \Psi - m \,e^\phi\, \bar{\Psi}\Psi\right),
\label{psieffective}
\end{equation}
where $\slashed D = \gamma^M(\partial_M + \frac{1}{4} \omega_{\underline{a}\underline{b}M}\gamma^{\underline{a}\underline{b}} - i q L^{-1} A_M)$, $\omega^{\underline{a}\underline{b}}$ is the spin-connection, and $\bar{\Psi} \equiv \Psi^\dagger \gamma^{\underline{t}}$. Underlined indices refer to  the local tangent space. Note that this action is part of a larger interacting theory resulting from the reduction of IIB supergravity, where the mass term appears with a dilatonic coupling as a consequence of this reduction.\footnote{Recently fermionic correlators were examined in  Einstein-Maxwell-dilaton gravity in \cite{Iizuka:2011hg} albeit without the dilaton coupling in the mass term.} In the specific reduced IIB  action the masses $m$ and charges $q$ take on a discrete set of values. Here we find it convenient to examine the behaviour for arbitrary $m$ and $q$, leaving the special top-down cases for future investigation.

Given this set-up it is a straightforward to analyze the Dirac equation resulting from \eqref{psieffective} in the background of the spinning D1-brane geometry. The fermionic correlators are obtained by a suitable modification of the prescription given in \cite{Iqbal:2009fd} to account for the fact that we have a geometry that is not asymptotically AdS. However, a simple trick allows us to reduce the computation to a previously solved problem and moreover provides a useful perspective. 
Therefore we will now take a detour into the mysterious world of supergravity truncations.

\section{Supergravity truncations}

It has long been known that the dilatonic branes geometries are conformal to AdS \cite{Boonstra:1998mp}. It is for instance clear that \eqref{ds3} is asymptotically conformal to AdS$_3$. What is more surprising is that this geometry has a hidden AdS$_4$ origin. For one, it can immediately  be oxidised to a Reissner-Nordstrom black hole in AdS$_4$ using the ansatz  (the gauge field lifts trivially)
\begin{equation}
ds_4^2 = \frac{ds_3^2}{r^2} +r^2 dy^2 = e^{2\phi}ds_3^2 +e^{-2\phi} dy^2
\label{dilatonlift}
\end{equation}
where $y$ denotes lift direction. This metric, in turn, solves field equations of the Einstein-Maxwell system with negative cosmological constant, 
\begin{equation}\label{s4}
S_4 = \frac{1}{16\pi\,G_4}\int  d^3x\, dy\, \sqrt{-g_4} \left(R(g_4)  - F^2 +  \frac{6}{L^2}\right).
\end{equation}
which is a well-known consistent truncation of 11D supergravity on ${\bf S}^7$. The action \eqref{s4} reduces to \eqref{s3} using the dilatonic ansatz \eqref{dilatonlift} and relating $G_4 = 2\pi R_y\, G_3$.

\begin{figure}[]
\includegraphics[width=0.6\columnwidth]{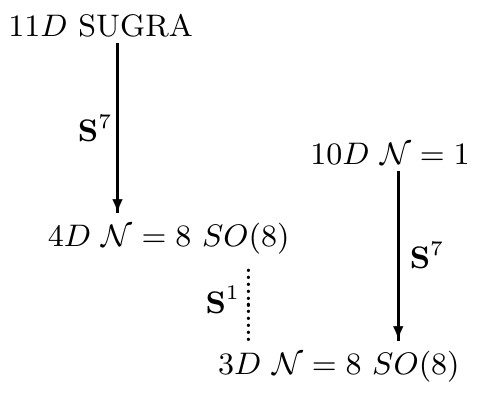}
\caption{\label{sugradiagram}Schematic hierarchy of supergravity truncations}
\end{figure}

Connections between these two systems have been observed previously \cite{Peet:1998wn, Mateos:2007vn, Kanitscheider:2009as, David:2009np,David:2010qc}. Here we go further and illuminate the origin of these connections as part of a more general story. The basic picture is presented in {\bf Fig.} \ref{sugradiagram}. 
Starting with $11$D  supergravity we may reduce  on ${\bf S}^7$ and perform a consistent truncation to the well known $\mathcal{N} = 8$ gauged $SO(8)$  supergravity \cite{deWit:1986iy}. The massless modes one retains are the graviton transforming in the singlet of $SO(8)$, gravitino in the ${\bf 8_s}$, while the fermions transform as ${\bf 56_s}$ and the scalars fill out ${\bf 35_s}$ and ${\bf 35_c}$. 

Likewise for Type IIA/B supergravity in $10$D where only a dilaton and 3-form field strength are retained, there remarkably is also a consistent truncation when reduced on ${\bf S}^7$ \cite{Cvetic:2000dm} to $\mathcal{N}=8$ gauged $SO(8)$ supergravity in $3$D. The  spectrum comprises of the graviton, a $SO(8)$ gauge field, 36 real scalars, transforming as  ${\bf 35_s}$ and a singlet of $SO(8)$, gravitinos transforming as ${\bf 8_s}$ and fermions in the ${\bf 56_s}$.

We have now at hand two different gauged supergravity theories with $SO(8)$ gauge group. For 4D solutions where no pseudo-scalars are sourced (the ${\bf 35_c}$), the resulting effective bosonic theories are related through an ${\bf S}^1$ reduction at the massless, nonlinear level. Whilst this is not a consistent truncation, it does allow us to relate a large class of solutions without pseudo-scalars, including the situation of current interest. 

Given this observation it follows abstractly that the Dirac equation arising from \eqref{psieffective} maps directly into the fermion wave equation on Reissner-Nordstrom AdS$_4$ background studied in \cite{Liu:2009dm, Faulkner:2009wj}. The map arises from simply KK-reducing on a spatial direction of this AdS$_4$, which amounts to restricting to the zero-momentum sector in the $y$-direction.\footnote{In addition one also projects the four-component spinor equation at $k_y=0$ into a pair of independent equations for two-component spinors.  One equation matches precisely that which arises from \eqref{psieffective}, whilst the other matches with $k_x \to -k_x$. } The spectral function we are after therefore is simply a slice of the spectral function for probe fermionic operators in the $2$d world-volume theory of M2-branes.

\section{Spectral functions}

\subsection{Numerical results at $q=3$, $m=0$, $T=0$}
Fermion Green's functions in theories with asymptotically AdS dual geometries are readily obtained using standard holographic techniques \cite{Iqbal:2009fd}. In Appendix \ref{s:appendix} we present the method adapted to the D1 geometry \eqref{ds3}.
\begin{figure}[]
\includegraphics[width=0.95\columnwidth]{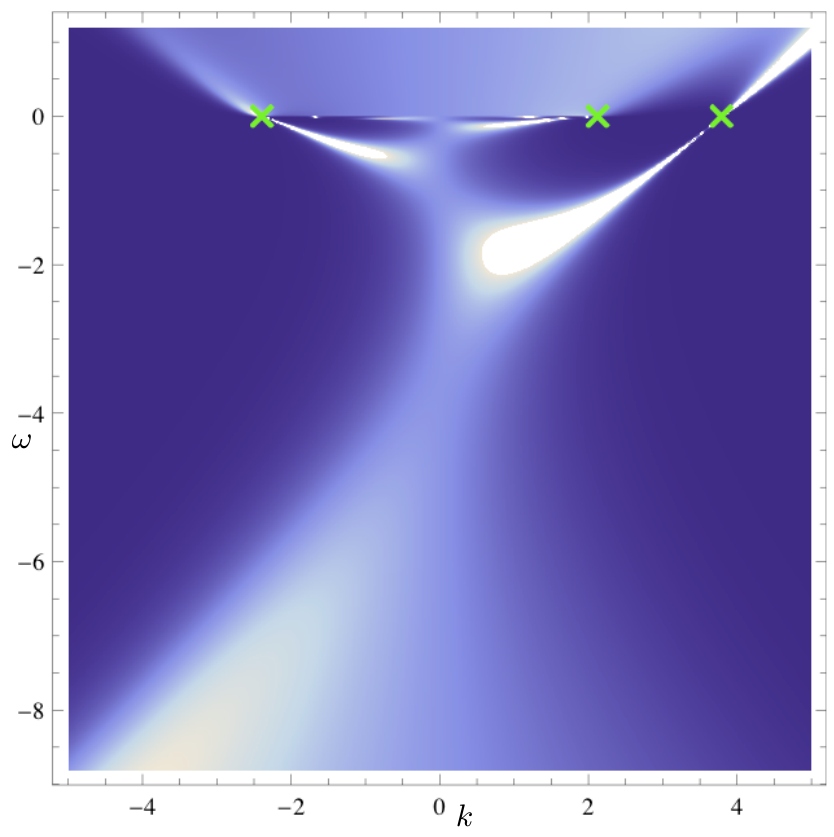}
\caption{\label{plot:spec}Density plot for the spectral function $\text{Im}\,G_R(\omega,k)$ of a fermionic probe of SYM$_2$ with $mL=0$, $q=3$ at chemical potential $ \mu L = Q = \sqrt{3}$ and $T=0$. The crosses indicate positions of normalisable, in-falling fermion modes at $\omega=0$. }
\end{figure}
Here, as a concrete example, we present numerical results for the case $q=3$, $mL=0$ at $T=0$. The  spectral function is presented in {\bf Fig.} \ref{plot:spec}, exhibiting features indicative of a Fermi surface. As expected these results match a $k_y=0$ slice of the M2 brane theory studied in \cite{Liu:2009dm}.

\subsection{Characterising the Fermi points}
Semi-analytic expressions for the $T=0$ correlators can be obtained for small frequencies, $\omega \simeq 0$, using a matched asymptotic expansion \cite{Faulkner:2009wj}. For the M2-brane geometry this relies on the fact that the geometry \eqref{dilatonlift} interpolates between a near horizon $\text{AdS}_2 \times {\mathbb R}^2$ and an asymptotically $\text{AdS}_4$ geometry. These results may be dimensionally reduced to yield expressions for Green's functions of the D1 system. Indeed, the analogous matching calculation for the D1 system yields identical results, as its quantum critical behaviour in the IR is determined by an $\text{AdS}_2 \times {\mathbb R}$ geometry consistent with the semi-holographic picture of \cite{Faulkner:2010tq}. 

The location of each peak is determined by the existence of a normalisable, in-falling fermion mode at $\omega=0$. These peaks are indicated by green crosses in the example of {\bf Fig.} \ref{plot:spec}. In the vicinity of each such peak we obtain,
\begin{equation}\label{greenspeak}
G_R(\omega, k) \simeq \frac{h_1}{k_\perp  - \frac{1}{v_F}\omega - \tilde{h}_2 \mathcal{G}_R(\omega,k_F)}
\end{equation}
where $k_F$ indicates the momentum of the peak in question, $k_\perp = k-k_F$. $h_1$, $\tilde{h}_2$ and $v_F$ are real and can be determined numerically. $\mathcal{G}_R(\omega,k)$ is the retarded Green's function computed in the near horizon $\text{AdS}_2 \times {\mathbb R}$:
\begin{equation}\label{innergreen}
\mathcal{G}_R(\omega,k)  = g(k)\,\omega^{2\nu_k}, \quad \nu_k \equiv \frac{1}{\sqrt{12}}\, \sqrt{2\,(k^2+m^2)L^2 - q^2} \,.
\end{equation}
The momentum dependent factor $g(k)$ can be found in \cite{Faulkner:2009wj} and  our result matches that of \cite{Faulkner:2011tm} (for the quantity $G_2$ with $k_y=0$).

\subsection{Peak broadening}
The type of quasi-particle behaviour depends on $\nu_{k_F}$ and there is no stable quasi-particle associated with peaks satisfying $2\nu_{k_F}<1$ \cite{Faulkner:2011tm}. When  $2\nu_{k_F}>1$, the magnitude of the AdS$_2$ contribution \eqref{innergreen} is subdominant to $\omega$ and in the vicinity of the pole at $\omega = k_\perp = 0$ we obtain,
\begin{equation}\label{nearpeak}
G_R(\omega, k) \propto \frac{1}{\omega - v_F k_\perp + i \gamma }, \quad \gamma \sim \omega^{2\nu_{k_F}} \sim k_\perp^{2\nu_{k_F}}.
\end{equation}
We have retained the leading imaginary contribution in the denominator, $\gamma$, which gives a width to the peak away from $k_\perp = 0$ and a finite quasi-particle relaxation time. 

The Green's function \eqref{nearpeak} is the same as a slice of the M2 theory, in particular, the quasi-particle width is the same. In contrast, the restricted phase space in the TL model leads to zero-width quasi-particle excitations. However, the 1d NLL models -- which modify TL by an $O(k^2)$ correction to the single-particle dispersion relation and density-density interaction -- exhibit Green's functions of the form \eqref{nearpeak} near the pole, with width $\gamma_{\text{\tiny NLL}}\sim k_\perp^{8}$ \cite{Khodas:2007uq}. The TL and NLL models also displays non-trivial power-law behaviour in the real-part of the Green's function, while \eqref{nearpeak} resembles a free-fermion-like linear dispersion near $k_\perp \simeq 0$.   Thus the  NLL models demonstrate that the differences between physics of Fermi points and higher dimensional Fermi surfaces need not be drastic, as  indeed we find here.

Of course, neither the TL nor the NLL models approximate the D1 system.\footnote{It is however amusing to note that excitations about a $\nu_{k_F} = 4$ Fermi point in the D1 system resemble the qualitative behaviour of relaxation via three particle scattering, the decay mode responsible for $\gamma_{\text{\tiny NLL}}$.} We simply wish to illustrate that it is possible to obtain a non-zero quasi-particle width by relaxing the approximation of a linear dispersion relation in 1d. This goes some way to resolving how the 1d and 2d systems of interest here may be connected microscopically.

\section{Discussion}

We studied a holographic model of a fermionic quantum liquid with a concrete dual QFT: $SU(N)$ SYM$_2$ with 16 supercharges.  Thus our holographic results provide in-roads to a system which may be tackled in other ways, such as lattice field theory. In the interest of phenomenology we focused on generic probes of this system; we leave supergravity specific probes for future study. It would also be interesting to study condensation of the adjoint scalars.

Where possible we confirm exact agreement between the direct holographic calculations in the D1 geometry and dimensional reduction of existing results with  AdS$_4$ asymptotics. The spectral functions for SYM$_2$ indicate the existence of quasi-particle peaks indicative of a Fermi surface, whose shape is controlled by the IR geometry and  exhibit a range of behaviours depending on $\nu_{k_F}$. 

The D1 spectral functions presented here differ from those of the TL model. The spectral support has a finite width about the quasi-particle peak (see e.g., {\bf Fig.} \ref{plot:spec}) indicative of nonlinear effects broadening the Fermi peak and is further consistent with lack of conformal invariance of SYM$_2$.

Of course, the TL model is idealised since the single particle spectrum is linearised about the Fermi points. Incorporating non-linearity in the dispersion enables the redistribution of quasi-particle momenta, leading to finite quasi-particle lifetimes \cite{Khodas:2007uq,Imambekov:2009fk}. In the holographic model, rather than having explicit modification of the fermionic spectrum, we find that the effects of the charged black hole background provide a non-linearisation of the spectrum. 

A curious feature of our analysis is the connection we have drawn between the SYM$_2$ and the M2-brane theories.  Whether this is an artefact of supergravity or holds deeper significance remains to be better understood. For the current discussion we note that it is rather amusing to realise that the analysis of fermionic correllators in the M2-brane theories can be related to a model in 1d, reminiscent of old ideas attributing strange metal phase of high $T_c$ superconductors to Luttinger liquid-like dynamics.

\begin{acknowledgements}
We thank J. Gauntlett, S. Gentle, V. Hubeny, H. Liu, J. McGreevy and S. Minwalla for helpful discussions. MR is supported in part by an STFC rolling grant. BW is supported by a Royal Commission for the Exhibition of 1851 Research Fellowship. 
\end{acknowledgements}

\appendix
\section{Green's functions in 3d}\label{s:appendix}
While we have emphasised the connection between the 3d and 4d effective actions in the main text, retarded Green's functions for $\mathcal{O}_\Psi$ may be obtained without invoking this using standard holographic techniques. We introduce $\psi  =  r^2 f(r)^\frac{1}{4} \Psi$ and pick a plane wave basis, $\psi(r,x,t) = e^{ikx-i\omega t} \psi_{\omega,k}(r)$. We further define projections under $\gamma^{\pm}  = \frac{1}{2}(1\pm\gamma^{\underline{r}})$ as $\psi^{\pm}_{\omega,k}(r) \equiv \gamma^{\pm} \psi_{\omega,k}(r).$ We supplement \eqref{psieffective} with the boundary term, $S_\partial = \int dt dx \sqrt{-h} \,i \bar{\Psi}^+ \Psi^-|_{r\to \infty}$ where $h$ is the induced metric on the boundary. The variation of the total fermion action $S_f = S_\Psi + S_\partial$, when on-shell is a boundary integral by virtue of the equations of motion:
\begin{equation}
\delta S_f \big|_{\text{on-shell}} = \lim_{r\to\infty} i\int dtdx \, \left(\psi^- \delta \bar{\psi}^+ -\bar{\psi}^- \delta \psi^+ \right)
\end{equation}
Counterterms are needed if $|mL|>\frac{1}{2}$, and for $|mL|=\frac{1}{2}$ there is a logarithm in the boundary expansion. For convenience we will focus here on $0\leq mL<\frac{1}{2}$. Retarded Green's functions $G_R(\omega,k)$ are given by imposing in-falling boundary conditions at the horizon, and reading off data at the boundary,
$G_R(\omega,k) =  r^{2mL}v_{\omega,k}(r) |_{r \to \infty},$
where $v_{\omega,k}(r) \equiv i\frac{\psi_{\omega,k}^-(r)}{\psi_{\omega,k}^+(r)}.$
The integration in the bulk is most convenient for the ratio $v(r)$ (dropping subscripts), since its equation of motion resulting from \eqref{psieffective} is first order,
\begin{equation}\label{firstorder}
\left(\sqrt{f(r)}\partial_r +\frac{2mL}{r} \right)v(r) + (U_t+U_x)v(r)^2  + (U_t-U_x) =0
\end{equation}
where $U= \left( -r^{-2}f(r)^{-\frac{1}{2}} \left(\omega L + qQ\left(1- r^{-1}\right)\right), Lr^{-2} k\right)$. Hence the Green's function is easily obtained by integrating \eqref{firstorder} from $v(1)=i$ to the boundary where $v(r)|_{r\to\infty}  = r^{-2mL}\left(G_R(\omega,k) +O\left(\frac{1}{r}\right)\right)$.

\bibliographystyle{apsrev}

\end{document}